\pdfoutput=1
\documentclass[aps,prl,twocolumn,showpacs,superscriptaddress]{revtex4-1}

\usepackage[pdftex]{graphicx}
\usepackage[pdftex]{epsfig}
\usepackage{epstopdf}
\usepackage{color}
\usepackage{amssymb,amsmath}
\usepackage{graphicx}
\usepackage{dcolumn}
\usepackage{multirow}
\usepackage{hyperref}

\newcommand{\bk}{{\bf k}}				
\newcommand{\bq}{{\bf q}}				
\newcommand{\br}{{\bf r}}				

\begin{document}

\title{X-ray Thomson scattering without the Chihara decomposition}
\author{A.D. Baczewski}
\affiliation{Center for Computing Research, Sandia National Laboratories, Albuquerque NM 87185}
\author{L. Shulenburger}
\author{M.P. Desjarlais}
\author{S.B. Hansen}
\affiliation{Pulsed Power Sciences Center, Sandia National Laboratories, Albuquerque NM 87185}
\author{R.J. Magyar}
\affiliation{Center for Computing Research, Sandia National Laboratories, Albuquerque NM 87185}

\begin{abstract}
X-Ray Thomson Scattering (XRTS) is an important experimental technique used to measure the temperature, ionization state, structure, and 
density of warm dense matter (WDM). The fundamental property probed in these experiments is the electronic dynamic structure factor (DSF). In most models, this is decomposed into three terms [Chihara, J. Phys. F: Metal Phys. {\bf 17}, 295 (1987)] representing the response of tightly bound, loosely bound, and free electrons. Accompanying this decomposition is the classification of electrons as either bound or free, which is useful for gapped and cold systems but becomes increasingly questionable as temperatures and pressures increase into the WDM regime. In this work we provide unambiguous first principles calculations of the dynamic structure factor of warm dense beryllium, independent of the Chihara form, by treating bound and free states under a single formalism. 
The computational approach is real-time finite-temperature time-dependent density functional theory (TDDFT) being applied here for the first time to WDM. We compare results from TDDFT to Chihara-based calculations for experimentally relevant conditions in shock-compressed beryllium.
\end{abstract}

\maketitle

Warm dense matter (WDM) arises in many contexts ranging from planetary science\cite{Guillot1999,Fortney2009a,Kraus2010} to the implosion stage of inertial confinement fusion\cite{Matzen2005,Drake2009,Hu2014}. While there are no sharp pressure, temperature and density boundaries for the WDM regime, it is generally viewed as an intermediate state between a condensed phase and an ideal plasma where Fermi degeneracy is present, and the Coulomb coupling and thermal energy are comparable in magnitude\cite{Glenzer2009}. 

Experimental characterization of warm dense matter is challenging due to the difficulty of producing uniform samples at extreme conditions and developing diagnostic techniques that can provide accurate and independent measurements of these conditions for transient samples opaque to optical photons. X-ray Thomson Scattering (XRTS)\cite{Gregori2003,Glenzer2009}, one such diagnostic technique,  
exploits the scattering of hard coherent x-rays 
to directly probe the system's dynamic structure factor (DSF). 
Through the fluctuation-dissipation theorem, the DSF is related to the system's density-density response, and consequently XRTS provides direct insight into electron dynamics.

XRTS experiments have been performed on a variety of materials including beryllium\cite{Glenzer2007,Lee2009}, lithium\cite{Saiz2008}, carbon\cite{Brown2014}, CH shells\cite{Fletcher2014}, and aluminum\cite{Ma2013,Fletcher2015}.   With recent improvements in source brightness\cite{Fletcher2015} producing increasingly high resolution and high signal-to-noise data, the full DSF is expected to become routinely available. In anticipation of these advances, it is critical and timely to examine the theoretical constructs underpinning the interpretation of these experiments.

The most common model of XRTS experiments relies on an additive form of the DSF due to Chihara\cite{Chihara1987,Chihara2000}:
\begin{equation}
 S(\bq,\omega)= |f_I(\bq)+\rho(\bq)|^2 S_{ii}(\bq,\omega) + Z_f S_{ee}(\bq,\omega) + S_{bf}(\bq,\omega) \label{eq:chihara_decomposition}
\end{equation}
The DSF varies with momentum and energy transfers ($\bq$ and $\omega$) and is partitioned into 3 features that can be interpreted in terms of x-ray scattering processes. These include scattering from electrons bound to and adiabatically following
ions ($|f_I(\bq)+\rho(\bq)|^2 S_{ii}(\bq,\omega)$), from $Z_f$ free electrons per ion ($Z_f S_{ee}(\bq,\omega)$), and from bound electrons that are photo-ionized ($S_{bf}(\bq,\omega)$). While successfully applied to many systems, this model relies on numerous approximations and assumptions. Most critically, the electrons are separated into bound and free populations, a distinction that is often ambiguous in the WDM regime.
Each term in Eqn. \ref{eq:chihara_decomposition} is subject to different models potentially leading 
to under-constrained fits to experimental data\cite{Souza2014}. The ionic feature typically relies upon a decomposition into a product of an ion-ion structure factor, $S_{ii}(\bq,\omega)$,  and an average atomic form factor, $f_I(\bq)+\rho(\bq)$, with the first term describing the unscreened bound electrons and the latter the screening cloud, which must be treated carefully\cite{Plagemann2015}. However, recent work has focused on moving past this decomposition of the elastic peak\cite{Vorberger2015}.

In this work, we transcend the Chihara decomposition by explicitly simulating the real-time dynamics of warm dense matter using a finite temperature form of time-dependent density functional theory (TDDFT)\cite{Ullrich2011Book,Marques2012Book} and the Projector Augmented-Wave (PAW) formalism \cite{Blochl1994,Kresse1999}. In PAW, the all-electron Kohn-Sham orbitals (and their associated density) can be accessed via an explicit linear transformation on the smoother pseudo orbitals. By performing calculations with none of the core states frozen, we avoid making any assumptions between bound and free electrons, and treat all electrons similarly. 

We work with a real-time implementation of TDDFT in the Vienna Ab-Initio Simulation Package (VASP)\cite{Kresse1996,Kresse1996a,Kresse1999} (see Supplemental Material \footnote{See Supplemental Material, which includes Refs. \cite{Qian2006,Ojanpera2012,SaadBook,MahanBook,Weissker2010,Brown1970,Perdew1992,
GoriGiorgi2000,Ceperley1980,Ortiz1999}}) that provides a number of attractive features. Physically, higher-order response phenomena and Ehrenfest molecular dynamics are accessible in this framework.  Computationally, the orthogonalization bottleneck that limits standard DFT approaches is removed, as it is only explicitly required for the calculation of the initial state of the Kohn-Sham orbitals. This leads to excellent strong scaling\cite{Andrade2012,Schleife2014}, and we have observed near-perfect scaling up to 65,536 cores in our implementation.

We next outline the details of our TDDFT calculations, noting that we use Hartree atomic units  ($m_e = e^2 = \hbar = 1/4\pi\epsilon_0 =1$) unless otherwise indicated. Time-dependent quantities evaluated at $t=0$ are indicated by the addition of a subscript $0$ and the absence of a temporal argument. Fourier transformed quantities are indicated by a diacritic tilde. All calculations are spin unpolarized.  

The equation of motion in real-time TDDFT is the time-dependent Kohn-Sham (TDKS) equation:
\begin{equation}
 i\frac{\partial}{\partial t} \phi_{n,\bk}(\br,t)=\left(-\frac{\nabla^2}{2} + v_S\left[\rho \right](\br,t)\right) \phi_{n,\bk}(\br,t) \label{eq:tdks_eqn}
\end{equation} 
in which $v_S\left[\rho \right](\br,t) = v_{ext}(\br,t) + v_H\left[\rho\right](\br,t) + v_{xc}\left[\rho \right](\br,t)$. The orbitals, indexed by band and Bloch wave number, are such that their weighted sum produces the time-dependent density, $\rho(\br,t)$. The external potential, $v_{ext}$, includes contributions due to the Coulomb field of the bare nuclei as well as a model of the x-ray probe, $v_{probe}(\br,t)$, which is quiescent until $t=0$. $v_H$ and $v_{xc}$ are the Hartree and exchange-correlation potentials, with accurate and efficiently computable approximations to the latter being a central theoretical concern of TDDFT.

The initial conditions for the $\lbrace \phi_{n,\bk}(\br,t)\rbrace$ from Eqn. \ref{eq:tdks_eqn} are the self-consistent solution to a Kohn-Sham Mermin DFT calculation\cite{Mermin1965} at electron temperature $T_e$ in a supercell of volume $\Omega_{sc}$. The initial density is then:
\begin{equation}
 \rho_0(\br) = \sum \limits_{n,\bk} f_{n,\bk}(T_e) |\phi_{n,\bk,0}\left[T_e\right](\br)|^2 
\end{equation}
where $f_{n,\bk}$ is a composite weight consisting of 
the measure of the specific Bloch orbital weighting
and the Mermin weight encoding temperature dependence according to a Fermi-Dirac distribution. 
It is important to note the implicit dependence of these initial orbitals on the equilibrium electron temperature, $T_e$. 
Evolving these orbitals under Eqn. \ref{eq:tdks_eqn}, the time-dependent density becomes:
\begin{equation}
 \rho(\br,t) = \sum \limits_{n,\bk} f_{n,\bk}(T_e) |\phi_{n,\bk}(\br,t)|^2
\end{equation}
The weights are not time-evolved, as we do not expect them to change in the linear response regime\cite{GiulianiBook}. For stronger perturbations, additional formalism might be required for the weights. That the Mermin formalism is sensible within TDDFT in the linear response regime is supported by recent foundational work\cite{Pribram2015a}.

\begin{figure}[ht]
 \includegraphics[width=\columnwidth]{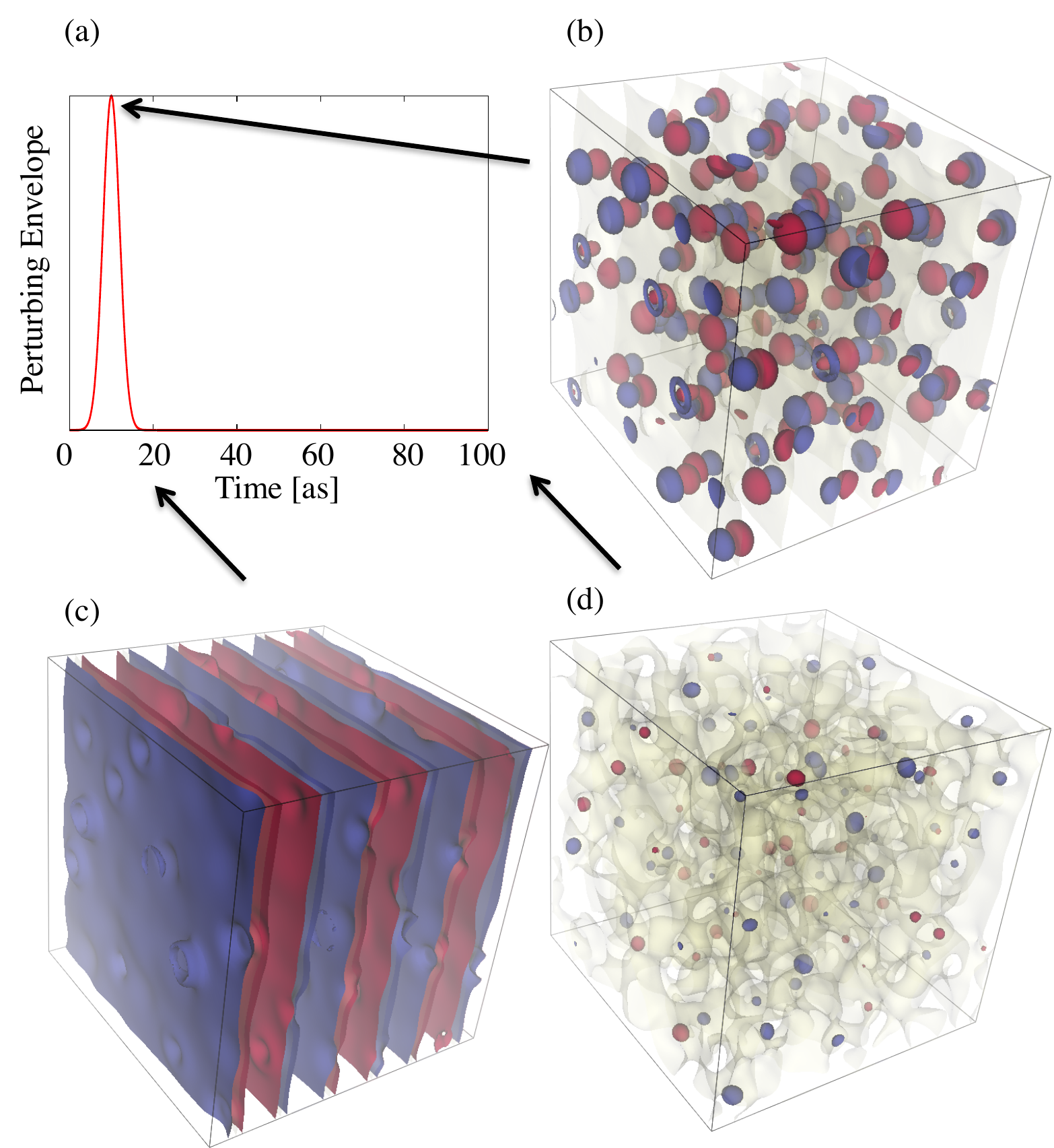}
 \caption{The density response of warm dense beryllium (density 5.5 g/cm$^3$ and $T_e=13$ eV, movie in Supplemental Material \cite{Note1}) due to (a) a perturbing potential with the illustrated envelope observed at times coinciding with (b) the peak of the perturbation, (c) the peak of the density response, and (d) the plasmons continuing to ring around the system after the perturbation has ceased. Red (blue) isosurfaces bound volumes of charge accumulation (depletion) and yellow isosurfaces indicate the nodal surface. The amplitude of the perturbation is within the linear response regime, a discussion of which is in the Supplemental Material \cite{Note1}. \label{fig:exemplary_calculation}}
\end{figure}

The  action of a probe potential, $v_{probe}(\br,t)$, turned on at $t=0$ leads to a change in the time-dependent density. The linear density-density response function, $\chi_{\rho\rho}(\br,\br',t)$, encodes the relationship between these two quantities:
\begin{equation}
 \delta \rho(\br,t) = \int \limits_{0}^{\infty} d\tau \int \limits_{\Omega_{sc}} d\br' \chi_{\rho\rho}(\br,\br',\tau) \delta v_{ext}(\br',t-\tau)
\end{equation}
where we use the notation $\delta f(\br,t) = f(\br,t) - f_0(\br)$ for $\delta \rho(\br,t)$ and $\delta v_{ext}(\br,t)$. If we fix ionic positions at time $t=0$, which evolve slowly on the relevant attosecond timescales, then $\delta v_{ext}(\br,t) = v_{probe}(\br,t)$ and we can construct a probe potential that can be used to extract the DSF, similar to Sakko, et. al.\cite{Sakko2010}. The real-time density response to such a probe potential is shown in Fig. \ref{fig:exemplary_calculation}.

In principle, any sufficiently weak analytic probe potential will allow us to extract the response function.  For convenience, we choose $v_{probe}(\br,t) = v_0 e^{i\bq\cdot\br}f(t)$, where $f(t)$ is a Gaussian envelope and $v_0$ is related to the probe intensity. The Fourier transformed response function, $\tilde{\chi}_{\rho\rho}(\bq,-\bq,\omega) = \delta\tilde{\rho}(\bq,\omega)/v_0 \tilde{f}(\omega)$, is then related to the DSF through the fluctuation-dissipation theorem:
\begin{equation}
 S(\bq,\omega) = -\frac{1}{\pi}\frac{\text{Im}\left[\chi_{\rho\rho}(\bq,-\bq,\omega)\right]}{1-e^{-\omega/k_bT_e}}
\end{equation}
We are careful to note that Fourier transforms are normalized such that $\delta\tilde{\rho}(\bq,\omega)$ has units of inverse frequency.

As a proof of principle, we report calculations of the DSF for 3x-compressed beryllium consistent with the conditions reported in  \cite{Lee2009}. We consider the same momentum transfers as in \cite{Plagemann2012} to study a range of excitations spanning the collective regime to the beginning of non-collective regime. For convenience of presentation, each $\bq$ value is mapped onto an XRTS scattering angle, $\theta$, relative to the 2 \AA~probe wavelength in \cite{Lee2009} (see Supplemental Material \cite{Note1} for more information). Our results come from averaging the response of electronic densities generated from several static uncorrelated ionic configurations sampled from thermally equilibrated DFT-MD calculations. These calculations were performed on 32 and 64 atom supercells with a four electron beryllium PAW potential within the local density approximation (LDA), with electrons and ions thermostatted at $T = 13$ eV, and $\bk$-point sampling at $(\frac{1}{4},\frac{1}{4},\frac{1}{4})$, analogous to the Baldereschi mean-value point for cubic supercells. For these conditions a plane wave cutoff of $1400$ eV was required to converge the pressure to within $1\%$, and 576 (32 atom)/1152 (64 atom) Kohn-Sham orbitals were needed to represent the thermal occupation ($f_{n,\bk}(T_e) \ge 10^{-5}$). 

\begin{figure}[ht]
 \includegraphics[width=\columnwidth]{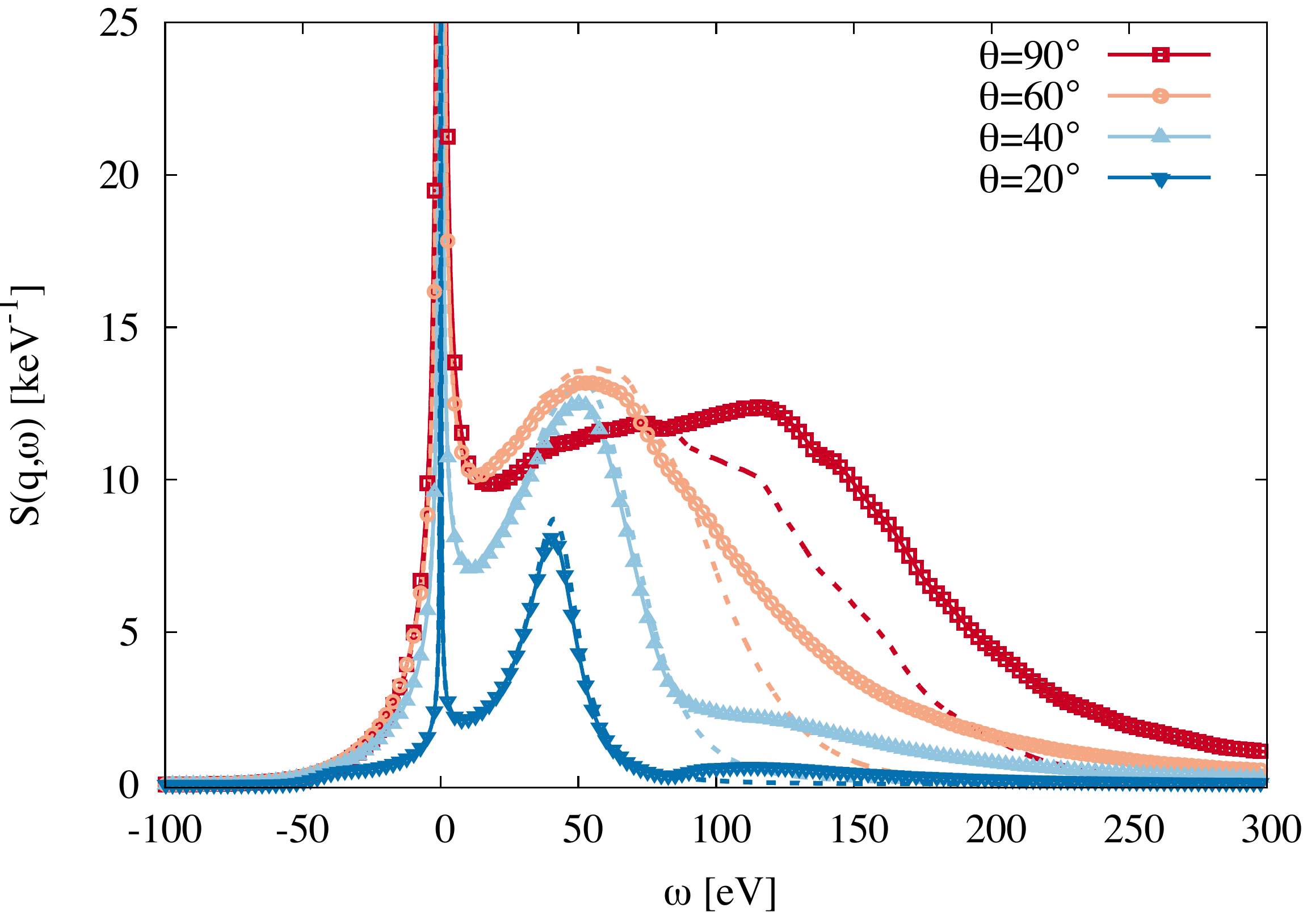}
 \caption{The DSF of warm dense beryllium (density 5.5 g/cm$^3$ and $T_e=13$ eV) at the scattering angles ($\theta$) considered in \cite{Plagemann2012} with (dashed lines) and without (marked lines) the frozen core approximation. All TDDFT calculations utilize the adiabatic LDA.\label{fig:beryllium_dsf}}
\end{figure}

\begin{figure}[ht]
 \includegraphics[width=\columnwidth]{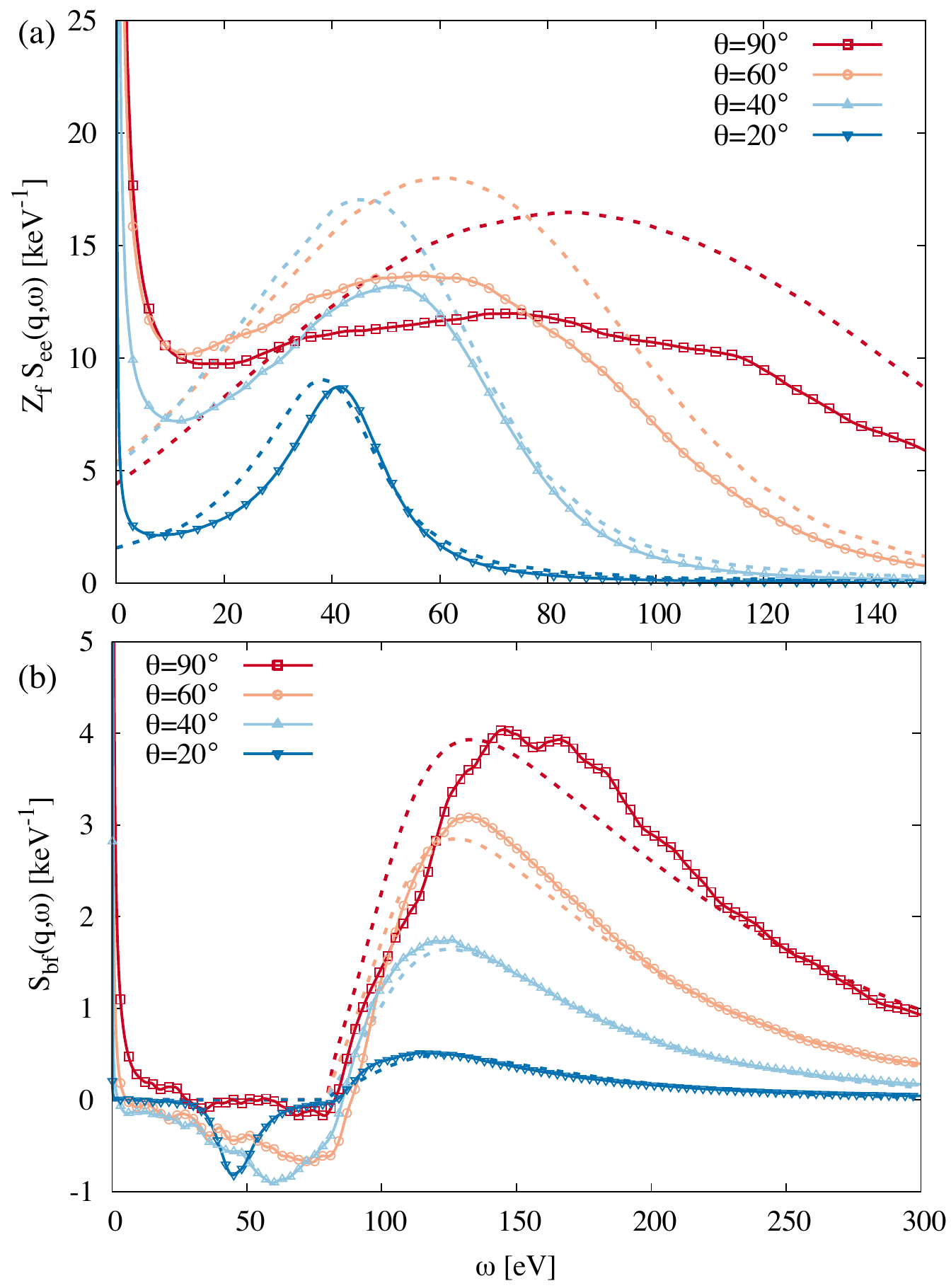}
 \caption{Comparison of the DSF of warm dense beryllium computed using TDDFT (marked lines) and individual terms in a Chihara decomposed theory (dashed lines). (a) Illustrates the two-electron TDDFT response compared to the MA-AICF method\cite{Plagemann2012} for $Z_f S_{ee}(\bq,\omega)$. (b) Illustrates the difference of the four- and two-electron TDDFT responses compared to an average atom treatment\cite{Johnson2012} of $S_{bf}(\bq,\omega)$. \label{fig:comparison}}
\end{figure}

Each sample configuration is used to seed a TDDFT calculation of the DSF, utilizing the same cutoff and number of orbitals. The initial Mermin electronic state is recomputed using a denser $\bk$-point sampling on a $3\times 4 \times 4$ (32 atom) or $2 \times 2\times 2$ (64 atom) Monkhorst-Pack grid. To assess the effect of the frozen core approximation (FCA), we consider electronic initial conditions and dynamics generated using both two and four electron beryllium PAW potentials. Results of our calculations are illustrated in Figs. \ref{fig:beryllium_dsf} and \ref{fig:comparison}. Details of the averaging procedure used to generate results, and information concerning the satisfaction of sum rules, can be found in the Supplemental Material \cite{Note1}.

Fig. \ref{fig:beryllium_dsf} directly compares the DSF computed with and without the FCA. While the PAW method still includes the proper all-electron density in aggregate, only the orbitals tied to the two outermost valence electrons are included in the time-evolved response within the FCA. This effectively removes the dynamics of the core states from the density response and the high energy shoulder above 80 eV in the DSF is removed. For the temperatures and densities being considered, this is roughly equivalent to partitioning the inner two and outer two electrons into bound and free groups in the Chihara picture, such that the two-electron response corresponds to $S_{ee}(\bq,\omega)$. However, there are important distinctions to keep in mind. First, in the four-electron calculation, all electrons are being treated identically, whereas it is typical to treat $S_{ee}(\bq,\omega)$ and $S_{bf}(\bq,\omega)$ using different levels of theory and without self-consistency in the Chihara framework. Second, even in the two-electron calculation, the response of the outer two electrons is still aware of the two frozen core states tied to each atom through their screening of the nuclear potential. Finally, these calculations are based upon explicit simulations of the real-time electron dynamics of a bulk supercell of warm dense beryllium rather than a phenomenological model of the response based upon a jellium plus average-atom picture. 

Based upon our observation that the two-electron response roughly corresponds to $S_{ee}(\bq,\omega)$, we can also extract a quantity akin to $S_{bf}(\bq,\omega)$ by differencing the four-electron and two-electron DSFs. The effective $S_{ee}(\bq,\omega)$ and $S_{bf}(\bq,\omega)$ computed within TDDFT are illustrated in Fig. \ref{fig:comparison}. Here we compare our TDDFT calculations to calculations done using state-of-the-art models for $S_{ee}(\bq,\omega)$ and $S_{bf}(\bq,\omega)$. The former is treated with an RPA-level model dielectric function with lifetime effects taken from four-electron DFT-MD calculations of the optical conductivity; the Mermin approximation-ab initio collision frequencies (MA-AICF) method in \cite{Plagemann2012}. The latter is calculated with the formalism developed in \cite{Johnson2012} and a quantum mechanical average-atom ion-sphere model with Slater exchange. As we are interested in studying energies relevant to the electronic response, we ignore $S_{ii}(\bq,\omega)$, though it is necessarily present in the Chihara-independent TDDFT calculation.

Examining the dispersion of the primary plasmon peak in Fig. \ref{fig:comparison}a, we see that TDDFT predicts a slight ($\sim$5 eV) blue shift relative to the MA-AICF calculation $\theta=20^\circ$ and $40^\circ$, whereas it predicts a stronger ($\sim$10 eV) red shift relative to MA-AICF at $\theta=60^\circ$ and $90^\circ$. We attribute this shift to exchange/correlation and band structure effects, not present in the MA-AICF dielectric function. Previously, comparisons of inelastic x-ray scattering spectra to RPA and LDA in cold free electron metals (Na and Al) indicate a similar trend in which both LDA and experiment are red shifted relative to RPA\cite{Cazzaniga2011}. However, these calculations also indicate that the addition of lifetime effects to the LDA are necessary to totally reconcile theory and experiment. While non-adiabatic exchange correlation kernels are available for energy domain linear response TDDFT, no such time-domain exchange correlation potentials are currently available for real-time TDDFT. As warm dense matter requires a large number of thermally occupied states such that energy domain TDDFT may become computationally prohibitive, warm dense matter may provide compelling motivation for the development and testing of these functionals.

Considering the bound-free shoulder in Fig. \ref{fig:comparison}b, we see that TDDFT is generally in good agreement with the average-atom model treatment of $S_{bf}(\bq,\omega)$ with some minor differences. Such average atom models agree well with real-space Green's function methods for cold solid beryllium \cite{Mattern2012}. We observe a trend opposite the free-free feature, in which there is a red shift of the TDDFT result at small angles, and a blue shift at large angles. We expect that LDA will do an increasingly poor job of describing the Compton scattering limit at large $\theta$ due to its well-established self-interaction error. Applying a functional with some fraction of Fock exchange should remedy this behavior and will be the subject of future investigations. Further, the TDDFT bound-free feature computed by differencing the two- and four-electron DSFs has a small negative dip below the 80 eV onset of the core feature. It is difficult to determine whether this is due to core-polarization suppressing the response of the valence electrons in the four-electron calculation, or potentially an artifact of the different pseudization procedures used to generate the two PAW potentials used for this comparison. This motivates further investigations of the PAW formalism applied to both real-time TDDFT and specifically, warm dense matter in which thermal effects will start to blur the line between core and valence electrons.

We presented a method for the direct calculation of the DSF for warm 
dense matter, independent of the Chihara decomposition, by applying real-time TDDFT to configurations drawn from thermal Mermin DFT-MD calculations. Comparison of our results with state-of-the-art models applied within the Chihara picture illustrates some subtle differences between the two, though it generally supports the use of the Chihara formalism as an inexpensive alternative to the very detailed and computationally intensive TDDFT calculations. We anticipate that TDDFT may provide a powerful discriminating tool for arbitrating disagreements between these more phenomenological theories and experiment, especially as experimental data becomes more highly resolved. Our framework enables future explorations of systems in which the partition between bound and free electrons is more ambiguous. It also provides a platform for studying the impact of recent foundational developments in DFT at non-zero temperature\cite{Li1985,Li1985a,Brown2013,Karasiev2014,Pribram-Jones2014,Pribram-Jones2014a,Pribram2015a}.

\section{Acknowledgments} \label{sec:acknowledgements} 
Calculations were completed on Chama and Skybridge at Sandia National Laboratories and Sequoia at Lawrence Livermore National Laboratory. We are grateful for discussions with 
Stephen Bond, 
Kieron Burke, 
Hardy Gross, 
Ryan Hatcher, 
Neepa Maitra, 
Thomas Mattsson,
Normand Modine, 
Jonathan Moussa, Kai-Uwe Plagemann, Aurora Pribram-Jones, Kenneth Rudinger, 
Travis Sjostrom, and 
Sam Trickey. 
A.D.B, L.S., M.P.D., and R.J.M. were supported by Sandia's Laboratory Directed Research and Development (LDRD) Project 165731. S.B.H. was supported by the U.S. Department of Energy, Office of Science Early Career Research Program, Office of Fusion Energy Sciences. Sandia National Laboratories is a multi-program laboratory managed and operated by Sandia Corporation, a wholly owned subsidiary of 
Lockheed Martin Corporation, for the U.S. Department of Energy's National Nuclear Security Administration under contract DE-AC04-94AL85000.

\clearpage
\widetext
\begin{center}
\textbf{\large Supplemental Materials: X-ray Thomson scattering without the Chihara decomposition}
\end{center}
\setcounter{equation}{0}
\setcounter{figure}{0}
\setcounter{table}{0}
\setcounter{section}{0}
\setcounter{page}{1}
\makeatletter
\renewcommand{\theequation}{S\arabic{equation}}
\renewcommand{\thefigure}{S\arabic{figure}}


\section{Implementation Details}

We have implemented real-time TDDFT using a plane wave basis and the Projector Augmented Wave (PAW) formalism \cite{Blochl1994,Kresse1999} within version 5.3.5 of VASP \cite{Kresse1996,Kresse1996a}. As in other implementations using ultrasoft pseudo-potentials \cite{Qian2006} or PAW \cite{Ojanpera2012} we chose a Crank-Nicolson (CN) time integration scheme to propagate the time-dependent Kohn-Sham (TDKS) equations numerically rather than chosing an integrator to achieve a high-order convergence in the local error. The unitarity of the discrete propagator was deemed an essential factor to guarantee the satisfaction of charge conservation and other sum rules associated with the dynamic structure factor (DSF) itself.  
At each timestep, the CN-discretized TDKS equations are solved using the Generalized Minimal Residual method (GMRES) \cite{SaadBook}. Given the perturbation on identity form of the discrete propagator we expect and observe rapid convergence in the number of iterations. 

As many details of our implementation have not been published elsewhere, we begin by demonstrating that it is robust. We first test stability of the time integration. Specifically, we illustrate that our implementation is not susceptible to any significant or uncontrolled errors, given the intrinsic nonlinearity of the TDKS equations and the use of an iterative solver on the linearized problem. To do so, we time propagate a Mermin thermal equilibrium set of orbitals in the absence of a probe potential or ionic motion. Here, the time-dependent potential should be constant in time, and any variations in the instantaneous total free energy or supercell charge are due to the accumulation of numerical errors. 

\begin{figure}[ht]
 \includegraphics[width=\textwidth]{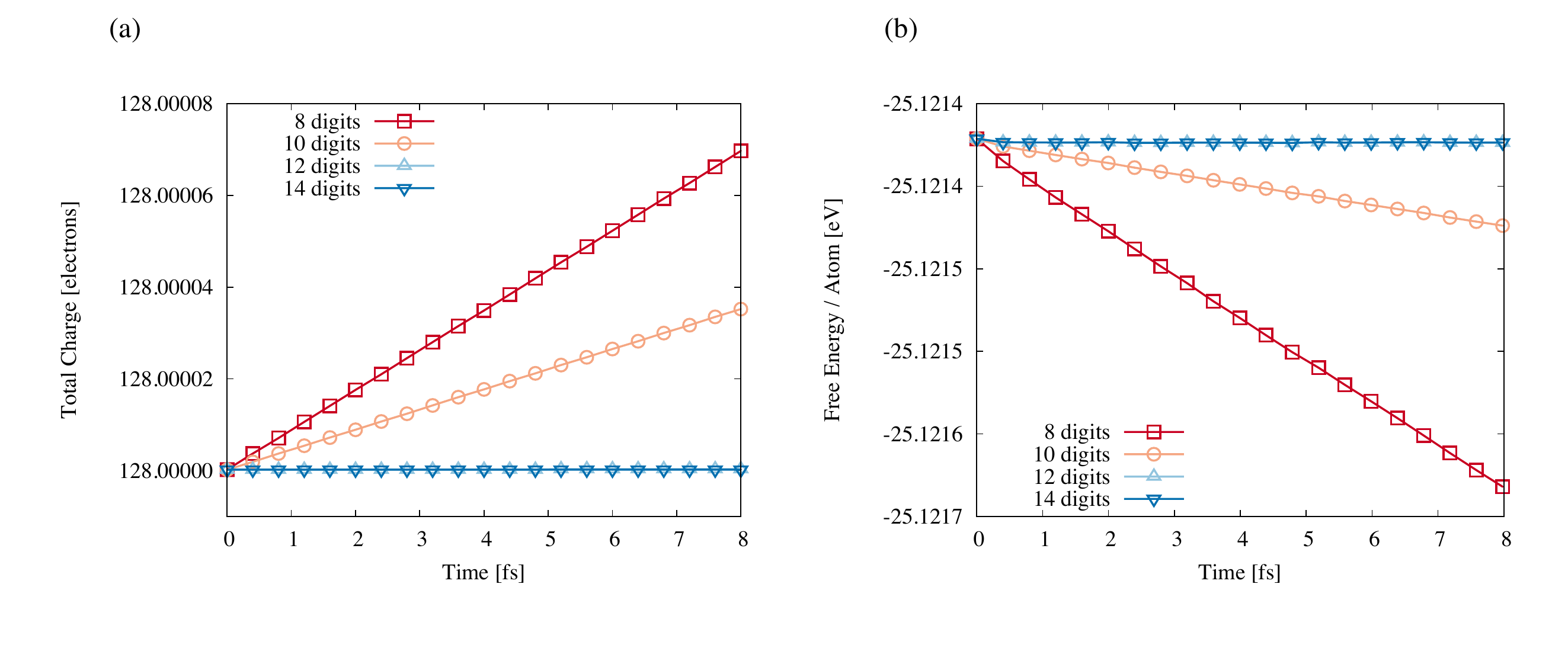}
 \caption{The (a) total charge of the supercell and (b) instantaneous total free energy per atom during the propagation of an effectively stationary Mermin state applying different relative tolerances for the convergence of GMRES. Noting that the temporal drift of both quantities in time is approximately linear, the reported drift is the slope of a linear regression. \label{fig:conservation}}
\end{figure}

Results are reported for 32 beryllium atoms under the warm dense conditions in the paper and using the adiabatic zero-temperature local density approximation throughout. The initial condition was generated from a Mermin DFT calculation done on a single atomic configuration drawn from a thermalized DFT-MD run with a four-electron PAW potential, 576 thermally occupied orbitals, and a plane wave cutoff of 1400 eV. The time propagation utilized the same plane wave cutoff and number of orbitals, and a time step ($\Delta t$) of 1 attoseconds (as) carried out for 8000 steps (8 fs). We varied the relative tolerance of the iterative solver, and kept the absolute tolerance fixed to be 2 orders of magnitude smaller. As the right hand side of the CN-discretized TDKS system is of order unit norm, the absolute tolerance is effectively irrelevant. The resultant free energy per atom and total charge density of the supercell are reported in Figure \ref{fig:conservation}.

Here, we see that the free energy drift per atom is $-2.6 \times 10^1 \mu$eV/(atom$\cdot$fs) and that the drift in the total charge of the unit cell is $8.7 \times 10^{-6}$ electrons/fs in the most permissive case (8 digits). In the least permissive case (14 digits), these quantities are $1 \times 10^{-3} \mu$eV/(atom$\cdot$fs) and $1.3 \times 10^{-11}$ electrons/fs, with the 12 digit case producing drifts of a similar order of magnitude. Going from 8 digits to 14 digits, the average CPU time per step varies linearly from 0.85 s/time step to 1.64 s/time step, i.e., convergence is exponential in the number of GMRES iterations. As the results are practically indistinguishable from 14 digits and the CPU time per time step is $20\%$ shorter, a relative tolerance of 12 digits was used in all subsequent calculations. Pertinent to the satisfaction of sum rules on $\delta \rho(\br,t)$, in all calculations in this work, we have verified that charge conservation is guaranteed to a relative accuracy of greater than 8 digits.

\begin{figure}[h]
 \includegraphics[width=\textwidth]{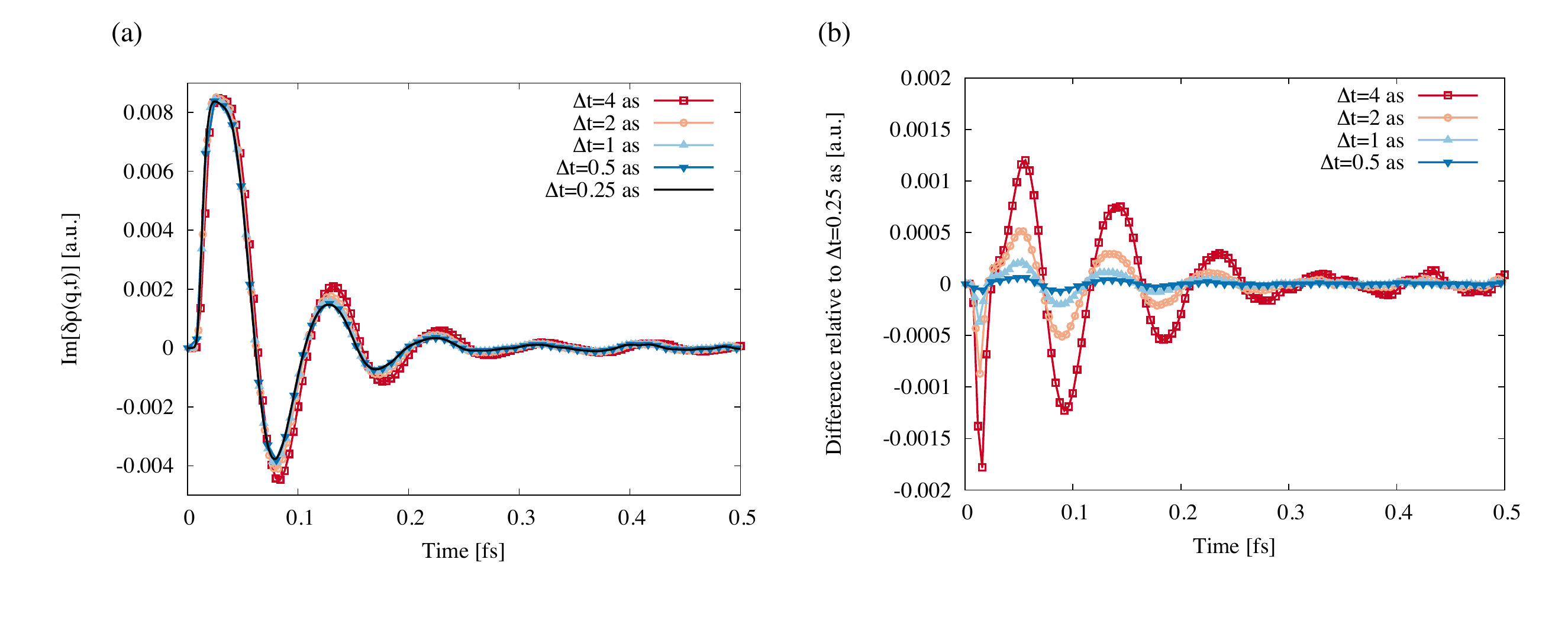}
 \caption{Convergence of $\delta\tilde{\rho}(\bq,t)$ in $\Delta t$. (a) Illustrates the $\text{Im}\left[\delta\tilde{\rho}(\bq,t)\right]$ for the sinusoidal component of the perturbation at different $\Delta t$ and (b) its instantaneous difference relative to $\Delta t = 0.25$ as. The relative $l_2$ distances from the $\Delta t=0.25$ as trajectory over a 2 fs interval are 5.7$\%$, 3.5$\%$, 1.9$\%$, and 0.8$\%$, in order of decreasing $\Delta t$. \label{fig:timestep_convergence}}
\end{figure}

We next demonstrate convergence of the integrated density response, the observable of interest, in $\Delta t$.  To do so, we consider the same 32 beryllium atom initial condition described above, this time applying a time-varying scalar perturbation of the form described in the main body of the paper. Here $v_{probe}(\br,t)=v_0 e^{i\bq\cdot \br}f(t)$ where $f(t)=\exp\left(-(t-t_d)^2/2t_w^2\right)/(\sqrt{2\pi} t_w)$ with $t_d=10$ as, $t_w=2$ as, $v_0=0.001$ eV$\cdot$fs, and $\bq=1.091\hat{x}$ \AA$^{-1}$. $\Delta t$ is varied from from $4$ as to $0.25$ as, and the associated convergence of the real-time density response is illustrated in Figure \ref{fig:timestep_convergence}.

From these results, it is evident that the density response exhibits first order convergence. Further, the convergence supports the decision that $\Delta t= 1$ as (attosecond) provides a reasonable balance between accuracy and time required per calculation for generating production results.

\begin{figure}[ht]
 \includegraphics[scale=0.5]{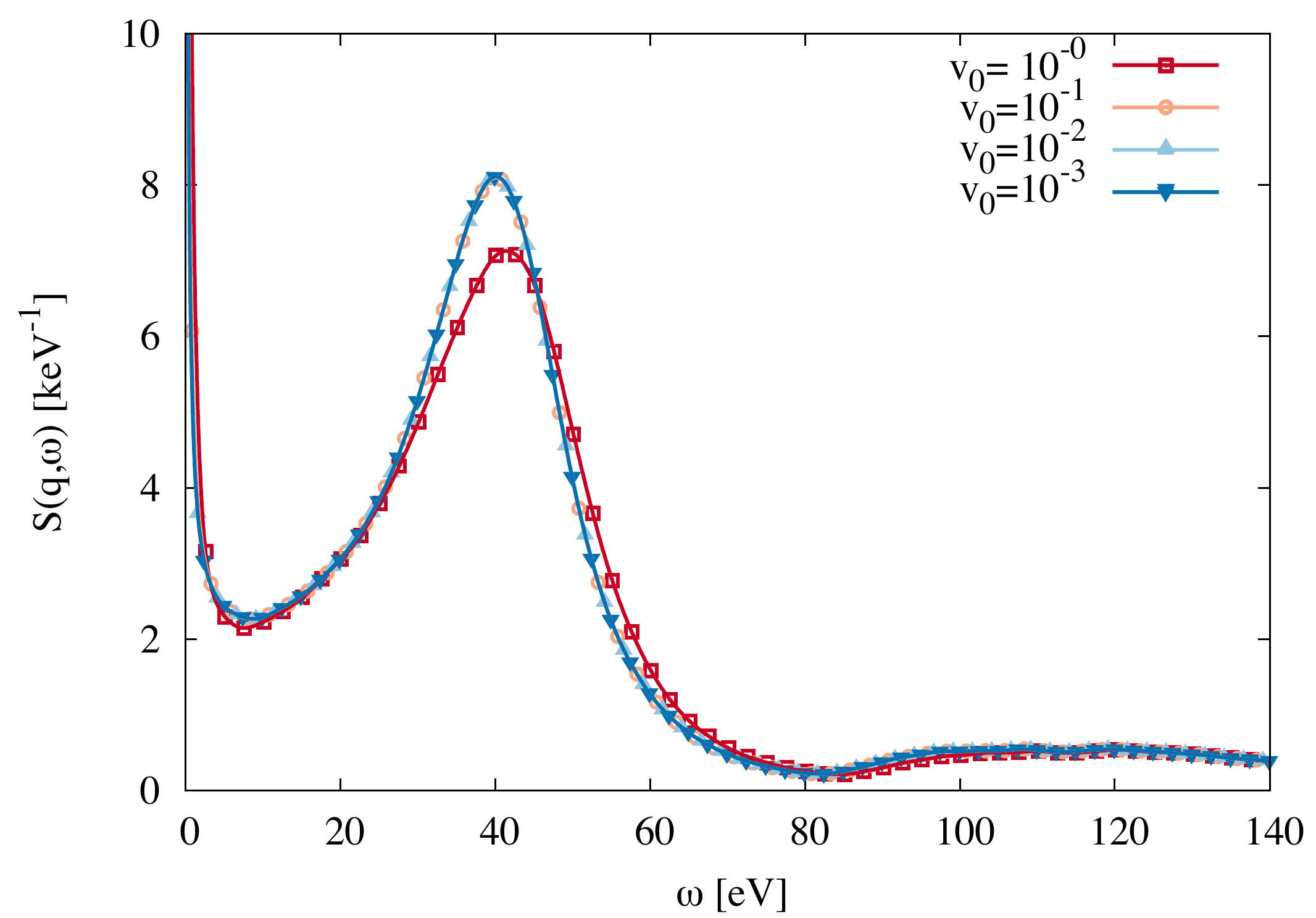}
 \caption{Variations in the DSF of warm dense beryllium as a function of probe amplitude ($v_0$ in eV$\cdot$fs). The spectra for small probe amplitude are indistinguishable indicating that we remain within the linear response regime. \label{fig:amplitude_variation}}
\end{figure}

Next, we consider the determination of parameters for $v_{probe}(\br,t)$. We chose $f(t)$ to take the form of a Gaussian envelope to ensure that the exciting potential and its response are approximately band- and time-limited. To this end, the pulse width ($t_w$), determines the bandwidth of the response and was chosen to ensure that modes of the density response with energies on the order of $100$s of eV are excited with appreciable amplitude. The delay, $t_d$, was chosen to ensure that the excitation is approximately quiescent at $t=0$ as. The remaining parameter, $v_0$, determines the effective intensity of the probe potential, and must be chosen to be large enough that the response is not dominated by numerical noise, yet small enough to remain in the linear response regime. Varying $v_0$ over 4 orders of magnitude from $1$ eV$\cdot$fs to $0.001$ eV$\cdot$fs, we do not observe numerical noise to be a problem at any value.  The post-processed DSF computed using these probe amplitudes for a single 32 atom configuration at $|\bq|= 1.091$\AA$^{-1} (\theta=20^\circ)$ are illustrated in Figure \ref{fig:amplitude_variation}. The results for $v_0$ ranging from $0.001$ eV$\cdot$fs to $0.1$ eV$\cdot$fs are indistinguishable, while the distortion of the results at $1$ eV$\cdot$fs indicate the onset of physics beyond linear response. That we can easily access this regime is one of the benefits of real-time TDDFT, though we do not explore this further in this work.

\newpage
\section{Generation of Initial Conditions}

Each TDDFT calculation requires a set of Kohn-Sham orbitals and occupancies as initial conditions. These are generated using DFT-MD as implemented in the standard version of the VASP software package \cite{Kresse1996a,Kresse1996,Kresse1999}. To assess the impact of the supercell shape and the underlying ionic positions on our DSF calculations, for each value of $\bq$ we run 2 separately thermalized DFT-MD trajectories on different supercells and draw 5 independent sets of initial conditions from each. Each set of initial conditions is used to seed independent TDDFT calculations of the DSF at a fixed $\bq$. 

Separate DFT-MD configurations are needed for each value of $\bq$ due to the requirements of the probe potential. For the DSF calculations, $v_{probe}(\br,t)$ must be commensurate with our supercell, and consequently any realizable value of $\bq$ must be in the reciprocal lattice of our supercell. To precisely specify the value of $\bq$ we work with tetragonal supercells in which the perturbing $\bq$ is directed along the c-axis. Then $\bq$ can be set by scaling the c-axis and the a-axes can be adjusted to ensure that the desired mass density is realized. Given the liquid-like ordering in the extreme conditions under investigation, we do not anticipate that this biases our results. Table \ref{table:geometries} gives the dimensions of the 8 supercells used in this study.

\begin{table}[ht]
 \caption{Parameters defining tetragonal supercell dimensions used in this study. 5 independent initial conditions were drawn from each row of this table to seed TDDFT calculations of the DSF at momentum transfer $\bq$. Each $\bq$ is equivalent to an XRTS scattering angle, $\theta$, through the relationship $|\bq|=4\pi \sin(\theta/2)/\lambda_0$ and $\lambda_0$ is the 2 \AA~probe wavelength from \cite{Lee2009}.  \label{table:geometries}}
 \centering
 \begin{tabular}{ c | c | c | c | c | c }
 \hline \hline
 a (\AA) & c (\AA) & \# of atoms & \# of bands & $|\bq|$ (\AA$^{-1}$) &  $\theta$ \\
 \hline 
 3.888 & 5.759 & 32 & 576 & 1.091 & $20^\circ$ \\
 3.858 & 5.848 & 32 & 576 & 2.149 & $40^\circ$ \\
 3.809 & 6.000 & 32 & 576 & 3.142 & $60^\circ$ \\
 3.923 & 5.657 & 32 & 576 & 4.443 & $90^\circ$ \\
 5.498 & 5.759 & 64 & 1152 & 1.091 & $20^\circ$ \\
 5.456 & 5.848 & 64 & 1152 & 2.149 & $40^\circ$ \\
 5.387 & 6.000 & 64 & 1152 & 3.142 & $60^\circ$ \\
 5.548 & 5.657 & 64 & 1152 & 4.443 & $90^\circ$ \\
 \end{tabular}
\end{table}

It is worth noting that all DFT-MD trajectories are generated with the full four-electron PAW potential. However, both the two-electron and four-electron TDDFT calculations are initialized from this same set of ionic configurations, with the Kohn-Sham orbitals and occupancies being recomputed for each set of fixed ionic positions in the two-electron case. This is done to assess the impact of the frozen core approximation on the DSF in isolation. 

\begin{figure}[ht]
 \includegraphics[scale=0.5]{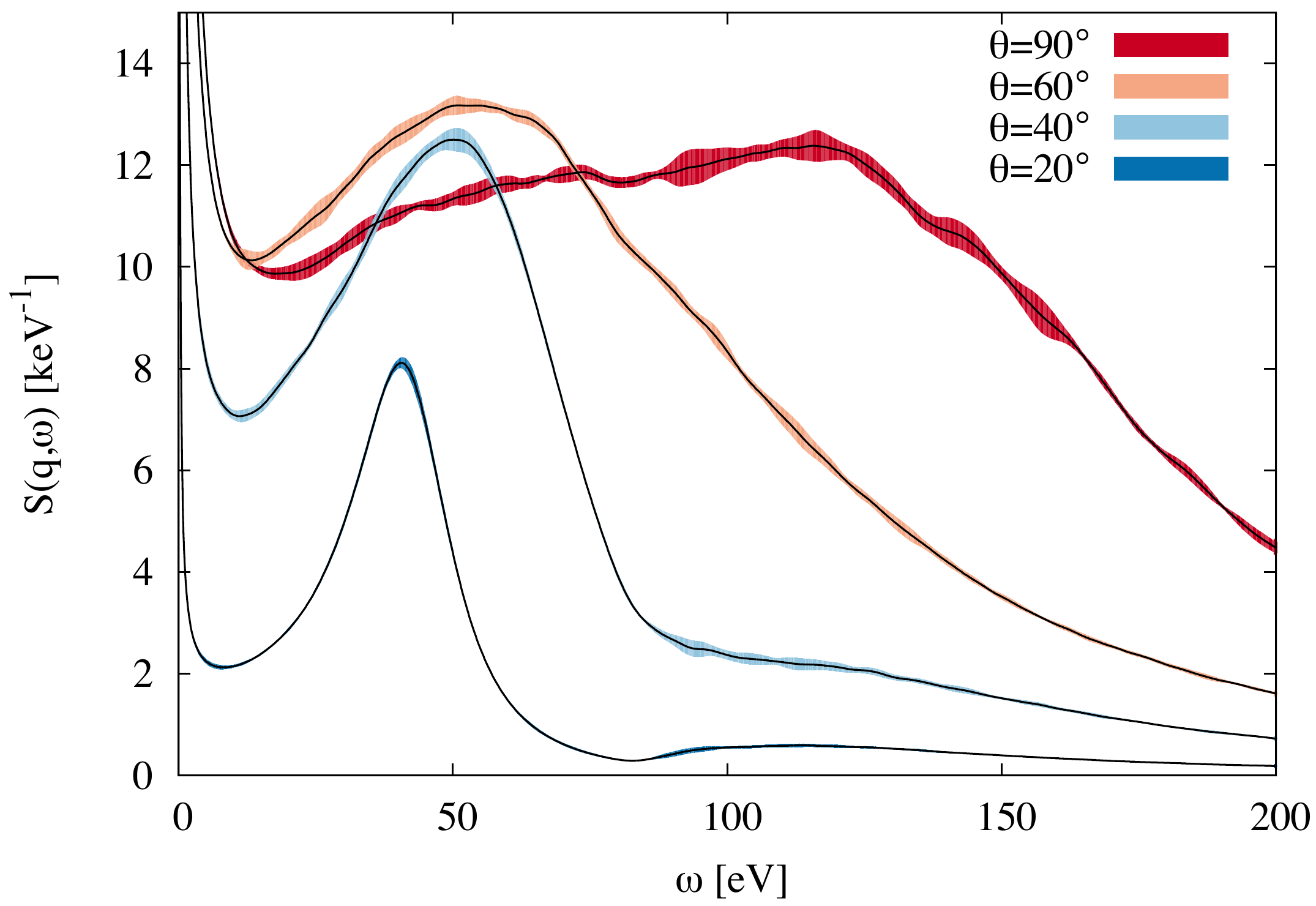}
 \caption{Jackknife error estimates of the DSF for warm dense beryllium at 4 different scattering angles. The colored shaded regions bound $\pm$4$\sigma$ about the mean, which is a solid black line. The $\pm$1$\sigma$ intervals are barely perceptible, and we simply present the mean values in the manuscript. \label{fig:dsf_variation}}
\end{figure}

All DSF results reported are averaged over 10 configurations sampled at each value of $\bq$. The sample size is necessarily small due to the significant computational resources required for each TDDFT calculation, with each production calculation being done on 1,152 cores. However, because the electronic density of warm dense beryllium is relatively uniform, we do not see much variability in the DSF from configuration to configuration. In an effort to quantify this variability, we apply jackknife resampling to estimate the variance and indicate single standard deviation intervals. For the scattering angles considered in the manuscript, we show the estimated error intervals for the DSF computed with four-electron PAW potentials in  Figure \ref{fig:dsf_variation}.

\section{Satisfaction of Sum Rules}

The units on the density response and DSF are such that the following form of the f-sum rule \cite{MahanBook} is satisfied:

\begin{equation}
 \int \limits_{-\infty}^{\infty} d\omega~\omega S(\bq,\omega) = \int \limits_{0}^{\infty} d\omega~\omega(1-e^{-\omega/k_B T_e}) S(\bq,\omega) = N_{e} \frac{|\bq|^2}{2}
\end{equation}
Here, $N_e$ is 2 for the two-electron PAW, and 4 for the four-electron PAW. Similar real-time calculations in the gas phase have reported satisfaction to within $5\%$ \cite{Sakko2010} and we report similar results. Forms of the DSF based upon model dielectric functions may or may not be consistent with various sum rules by construction. Our DSF is derived strictly from a time-evolved electronic density for which charge conservation is numerically guaranteed to high precision, and makes no assumptions about the form of the equivalent dielectric response. To this end, the primary source of error in our satisfaction of sum rules is due to numerical errors in the post-processing, e.g., the numerical evaluation of Fourier integrals with integrands which become increasingly oscillatory at higher energies, and thus more prone to small errors in the response. When checking sum rules produced using linear response TDDFT, it is common practice to fit the high energy tail of the DSF to force it to go to zero in a way that is consistent with the integrand in the above sum rule \cite{Weissker2010}. This is also critical for real-time simulations where small phase errors between the real and imaginary parts of the time domain density response are amplified at high energies relative to low energies when post-processing to generate the energy domain response. Rather than force the tail to fit some form, we simply impose a high energy cutoff on the integrand once the DSF has decayed to $< 1\%$ of its peak value. We simply seek to verify that the majority of the weight of our response is consistent with the sum rule, and could resort to fitting tails if necessary to improve agreement.

Applying this technique, we verify that we satisfy the f-sum rule for our four-electron data with a relative error of $-7\%$ ($\theta=20^{\circ}$), $-3\%$ ($\theta=40^{\circ}$), $-2\%$ ($\theta=60^{\circ}$), and $-4\%$ ($\theta=90^{\circ}$). In all cases, we underestimate the value of the f-sum rule, giving us confidence that a fit of the slowly-decaying high energy tail might be used to improve this agreement. Applying this same technique to the two-electron data we find relative errors of $8\%$ ($\theta=20^{\circ}$), $5\%$ ($\theta=40^{\circ}$), $0.4\%$ ($\theta=60^{\circ}$), and $-2\%$ ($\theta=90^{\circ}$). Here, we do not uniformly underestimate the sum rule as was the case with the four-electron data. The two-electron data decays much more abruptly at high energies, and does not need to represent the response of slowly-decaying core states, so we do not view this as a deficiency (i.e., fitting a tail would not have as strong of an impact here). However, this seems to point to the two-electron response as overestimating the free-free peak, consistent with the small negative dips in the bound-free data in Fig. 3b for $\theta=20^{\circ}$, $40^{\circ}$, and $60^{\circ}$ (the calculations for which the sum rule was over-estimated). Whether or not this can be improved with a different two-electron PAW may be an interesting topic of further study.

\begin{figure}[ht]
 \includegraphics[scale=0.5]{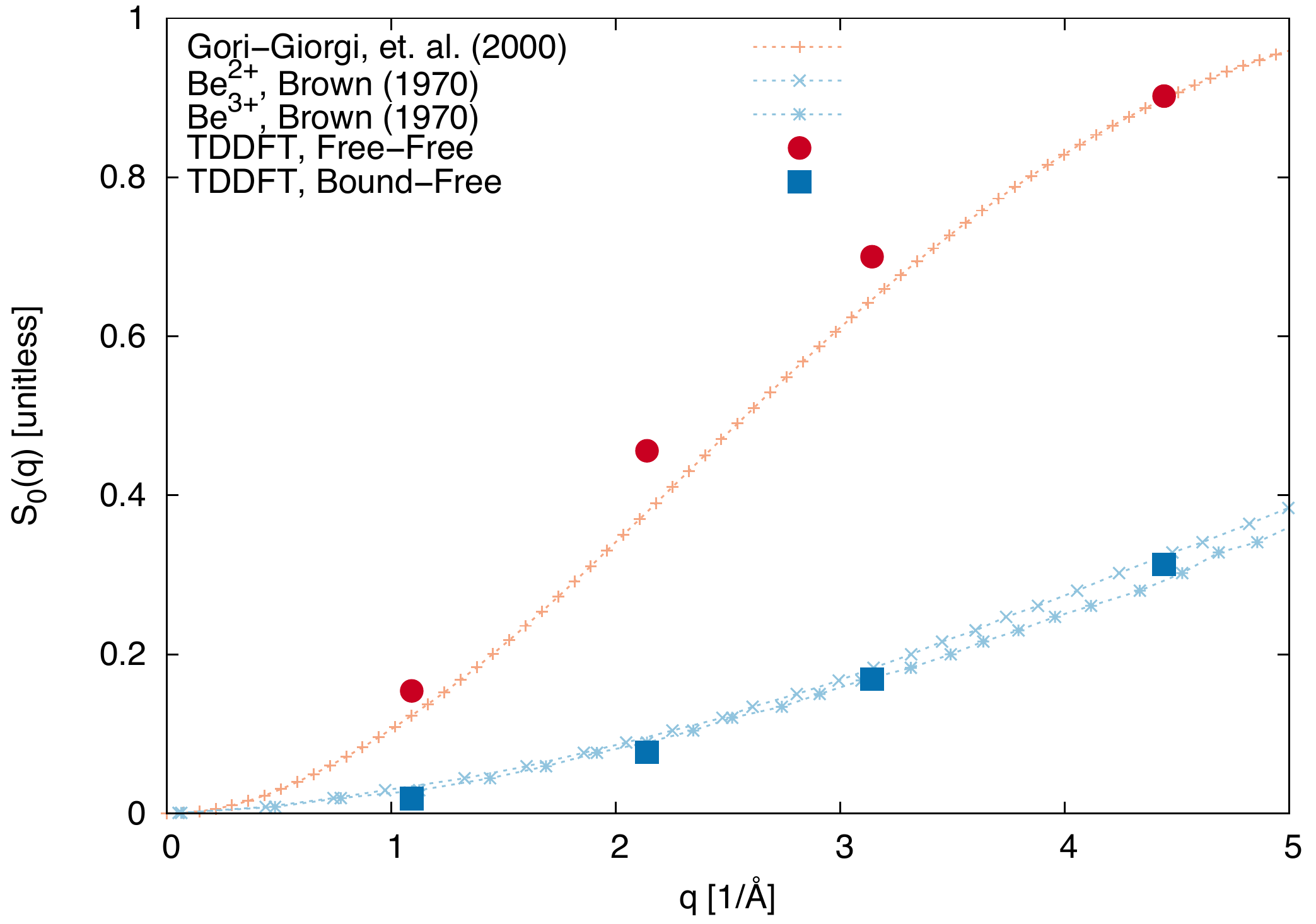}
 \caption{The static structure factor computed using Eqn. \ref{eq:static_structure} applied to the effective free-free and bound-free responses extracted from TDDFT. The bound-free structure factor is compared to that of doubly- and triply-ionized beryllium \cite{Brown1970}.\label{fig:static_structure_factor}. The free-free structure factor is compared to an analytic fit of QMC data for jellium with $r_s=1.3$ a.u \cite{GoriGiorgi2000}.}
\end{figure}

Perhaps a more interesting sum rule to study is the one that defines the static structure factor through the integral of the DSF:

\begin{equation}
 \int \limits_{-\infty}^{\infty} d\omega S(\bq,\omega) = N_e S_0(\bq) \label{eq:static_structure}
\end{equation}
Computing this integral for the effective free-free and bound-free data from TDDFT, we can extract structure factors that we can compare against physically intuitive models. Results are presented in Fig. \ref{fig:static_structure_factor}.

For the bound-free component we compare to results obtained using a configuration-interaction expansion by Brown for doubly- and triply-ionized  beryllium \cite{Brown1970}. For $\theta=60^\circ$ and $90^\circ$, the TDDFT gives results that are closer to doubly-ionized beryllium, consistent with physical intuition for the thermodynamic conditions under consideration. For $\theta=20^\circ$ and $40^\circ$, the TDDFT gives results that are closer to triply-ionized beryllium, though the difference between the doubly- and triply-ionized structure factors are smaller at these angles. We note that by excluding the negative difference data below 80 eV in our integration (evident in Fig. 3b), we can improve agreement with the doubly ionized curve at all angles. This indicates that the underestimation of the TDDFT structure factor may be due to differences between the free electron response for the two-electron and four-electron PAWs, which may or may not be physical. This highlights the importance of being able to self-consistently compute the four-electron response without the Chihara decomposition using our methodology.

For the free-free component (our frozen core result) we compare to results for the static structure factor of jellium with $r_s=1.3$ a.u., consistent with the experimentally determined free-electron density \cite{Lee2009}. In evaluating Eqn. \ref{eq:static_structure}, we remove the elastic peak in a region of width $\sim k_B T_e$ centered at $\omega=0$ and apply a simple linear interpolant between the DSF data on either side of the excluded region. We compare the resultant free-free static structure factor to an analytic fit to QMC data \cite{Ortiz1999} by Gori-Giorgi, et. al. \cite{GoriGiorgi2000}. The TDDFT free-free structure factor exhibits the same trend as the QMC fit, but is not expected to agree perfectly. We only expect qualitative agreement because the external potential experienced by the free electrons in warm dense beryllium is {\it not} a uniform neutralizing background as is the case in jellium.

This qualitative agreement between results from TDDFT and QMC stands in contrast to Vorberger and Gericke's recent work in which DFT-MD is used to compute a free electron density, from which the free-free static structure factor of warm dense beryllium is extracted \cite{Vorberger2015}. In their work, the static structure factor computed from DFT differs greatly from QMC, namely it does not go from one to zero for momentum transfers less than $2 k_F$ ($k_F = 2.88$ \AA$^{-1}$ for our conditions), but instead oscillates slightly about unity. The authors postulate that this is due to the mean field nature of the Kohn-Sham equations, and that this may be beyond the capability of DFT. Our results indicate that this basic physics is in fact within the grasp of TDDFT. To this end, it is worth noting that TDDFT has provided us with a convenient means of computing the static structure factor as an exact functional of the time-dependent density. In this case, we mean exact in the sense that if we are given a representation of the exact interacting time-dependent density, we can map it onto the exact DSF and the exact structure factor through Eqn. \ref{eq:static_structure}.

\end{document}